\documentclass[12pt]{iopart}

\usepackage{graphicx}

\def\be{\begin{equation}}
\def\ee{\end{equation}}
\def\bea{\begin{eqnarray}}
\def\eea{\end{eqnarray}}

\begin{document}

\title[String Gas Cosmology]{String Gas Cosmology after Planck}

\author{Robert H. Brandenberger}

\address{Physics Department, McGill University, 3600 University St., Montreal, QC, H3A 2T8, Canada}
\ead{rhb@physics.mcgill.ca}
\vspace{10pt}
\begin{indented}
\item[]April 2015
\end{indented}

\begin{abstract}

We review the status of String Gas Cosmology after the 2015 Planck
data release. String gas cosmology predicts an almost scale-invariant
spectrum of cosmological perturbations with a slight red tilt, like the
simplest inflationary models. It also predicts a scale-invariant spectrum
of gravitational waves with a slight blue tilt, unlike inflationary models
which predict a red tilt of the gravitational wave spectrum. String gas cosmology
yields two consistency relations which determine the tensor to scalar
ratio and the slope of the gravitational wave spectrum given the amplitude
and tilt of the scalar spectrum. We show that these consistency relations
are in good agreement with the Planck data. We discuss future observations
which will be able to differentiate between the predictions of inflation
and those of string gas cosmology.

\end{abstract}

\section{Introduction}

The recently released Planck results \cite{Planck} have further confirmed the
predictions of the six parameter $\Lambda$CDM cosmological model. Even
though these six parameters describe properties of the current universe, their
values are quite mysterious without going back to processes which happened
in the very early universe. In particular, the origin of the almost scale-invariant
spectrum of almost adiabatic curvature fluctuations with a slight red tilt which 
has now been confirmed by Planck and many other experiments has no
explanation in the context of late time cosmology.

Inflationary cosmology \cite{Guth} is the current paradigm for explaining the overall
homogeneity and spatial flatness of space and the origin of the observed
spectrum of fluctuations \cite{Mukh}. The inflationary scenario is based
on the potential energy of some scalar field yielding a period of (almost)
exponential expansion of space. In simple models of inflation, however, the
energy scale at which inflation takes place is of the order of $10^{16} {\rm GeV}$,
much closer to the Planck scale than to scales which have been explored
in particle physics experiments, and in fact even closer to the string 
scale \cite{GSW}. Hence, in order to truly understand the mechanism of
inflation, it appears to be necessary to tackle superstring cosmology.

Another reason for turning to string theory as a framework for inflationary
cosmology is the ``$\eta$ problem'' (see e.g. Section 27.2 in \cite{etaproblem}). 
In order not to produce 
a too large amplitude of fluctuations, the scalar field potential must be sufficiently
flat. For simple scalar field potentials, the slow-roll inflationary dynamics takes place
at field values larger than the Planck mass $m_{pl}$. This is also a field
range for which the slow-roll inflationary trajectory is a local attractor in
initial condition space \cite{initial}, even including linear metric fluctuations \cite{Hume}
\footnote{However, beyond linear order the story may be very different
\cite{Tanmay, Paul, Mark}. Also, in the case of small field inflation there is
an initial condition problem \cite{Piran}.}. In order to understand physics in this
field range, it seems necessary to embed inflation into an ultraviolet complete
theory of quantum gravity such as superstring theory. 

In most current approaches to superstring cosmology, the theory is treated
in an effective field theory limit in which scalar fields motivated by string 
physics are coupled to Einstein gravity or dilaton gravity (see e.g. \cite{reviews}
for recent reviews on inflation in this framework). However, this approach
misses one of the key symmetries of string theory, namely the ``T-duality''
symmetry \cite{Pol}. To illustrate one manifestation of this symmetry,
assume that space is toroidal. In this case, the perturbative spectrum of
string states constains both string momentum and winding modes.
The energy of the string momentum modes scales as $1/R$, where
$R$ is the radius of the torus, whereas the energy in the winding
modes increases linearly in $R$. Thus, the spectrum of string states
is unchanged under the transformation $R \rightarrow 1/R$ (in string units).
The string vertex operators are consistent with this symmetry, and if
we postulate that the symmetry extends to the non-perturbative level,
we predict the existence of D-branes \cite{Pol, Boehm}. 

Another feature of superstring theory which is missed in the effective
field theory description is the existence of an infinite tower of string
oscillatory modes which leads to a maximal temperature for a gas
of closed strings in thermal equilibrium, the {\it Hagedorn temperature}
\cite{Hagedorn}. It is self-evident that both the T-duality symmetry
of string theory and the existence of a limiting temperature should
play an important role in superstring cosmology.
The {\it string gas cosmology} (SGC) \cite{BV}  scenario which will
be reviewed here is an approach to superstring cosmology
which is based on the T-duality symmetry of string theory and string
thermodynamics \cite{Nivedita}. As was realized much later \cite{NBV},
string gas cosmology yields an alternative to inflation for explaining
the origin of the inhomogeneities and anisotropies which are
now mapped out by cosmological experiments such as the Planck
satelllite mission.

Some readers may be under the impression that observations
have proven that there was a period of inflation in the early universe.
However, all that the observations show is that there is some
mechanism which produces an almost scale-invariant spectrum
of nearly adiabatic fluctuations with a small red tilt on scales
which were larger than the Standard Big Bang horizon at early
times (times comparable to the time $t_{rec}$ of last scattering).
That such a spectrum yields acoustic oscillations in the angular
power spectrum of the cosmic microwave background (CMB)
and baryon acoustic oscillations in the galaxy power spectrum
was realized \cite{Sunyaev} more than a decade before
the development on the inflationary scenario. It is true that
cosmological inflation is the first model where such a spectrum
emerges from first principles, but it is not the only one (see e.g.
\cite{RHBrev1} for a recent review of alternatives). 

Alternatives include the {\it Pre-Big-Bang} and {\it Ekpyrotic} 
scenarios \cite{Ekp} (in which
a light entropic mode obtains a scale-invariant spectrum during
the contracting phase \cite{NewEkp}), the {\it conformal
universe} \cite{Rubakov} or {\it pseudo-conformal universe} 
\cite{Khoury} (in which a scale-invariant spectrum is generated
from an emergent phase by a moving Galileon field which induces
squeezing of the curvature fluctuations), the {\it matter bounce}
scenario \cite{Wands, Fabio} in which a scale-invariant spectrum
of curvature perturbations is generated from initial vacuum
fluctuations on scales which exit the Hubble radius during the
matter-dominated phase of contraction, and {\it string gas cosmology},
the topic of this review. In the same way that inflation may find an
embedding within superstring theory, it is also possible that
one of the alternatives to inflation ends up being realized in
string theory. For example, the Ekpyrotic scenario was initially
motivated from a stringy construction, heterotic M-theory \cite{Horava},
and there is a recent realization \cite{Costas} of the matter-bounce
scenario in which an S-brane originating from extra states
becoming massless at a T-dual point mediates the transition
from contraction to expansion. String gas cosmology is, obviously,
from the outset firmly rooted in superstring theory.
 
In the following we first present a review of string gas cosmology
(see e.g \cite{SGCreviews} for more detailed reviews). We then
confront the predictions of string gas cosmology with
recent observations, and we close this article with a discussion
of some major challenges facing the scenario.

\section{String Gas Cosmology}

Standard Big Bang Cosmology is based on coupling a gas of point
particles to a background space-time. ``String Gas Cosmology''
\cite{BV} (see also \cite{Perlt}) is a modest extension
of this setup which maintains the key new symmetries (T-duality)
and new degrees of freedom (string oscillatory and winding modes)
which distinguish string theory from point particle theories.
This is achieved by replacing the gas of point particles by a 
gas of strings. To be specific, we consider a theory of
closed superstrings, and we assume that the background space
is toroidal (extensions to toroidal orbifolds are considered
in \cite{BG1}). We also assume that the string coupling constant
is small such that the strings are the light degrees of
freedom (compared to branes - for an analysis of the role
of branes in string gas cosmology see \cite{Stephon}).

Following what is done in Standard Big Bang Cosmology, we assume
that matter is in thermal equilibrium. The infinite tower of
string oscillatory modes then leads immediately to a crucial 
difference between point particle cosmology and string cosmology:
the temperature of the gas of strings has a maximal temperature
$T_H$, the ``Hagedorn temperature'' \cite{Hagedorn} which is given
by the string scale. Let us start with a large dilute box of strings at
low temperatures. In this case almost all of the energy is in the
momentum modes whose energy increases as the box contracts.
This leads to a rising temperature, as in particle cosmology. However,
once the temperature approaches $T_H$, the energy density is high
enough to excite the string oscillatory modes. Further contraction of
the box will lead to a growing tower of oscillatory modes being excited
at approximately constant temperature. Once the box size decreases
below the string scale, the energy of the gas of strings will drift into
the winding modes which become less energetic as the universe
contracts, leading to a deceasing temperature. Figure 1 \cite{BV} 
shows how the temperature of a gas of heterotic superstrings in a
box of radius $R$ varies as the radius changes. The vertical
axis is temperature, the horizontal axis the radius of the box in units
of the string length. The two different curves corresponding to 
different total amounts of entropy - the larger the entropy the
more extended the {\it Hagedorn phase}, the phase where $T$ is
close to $T_H$.

\begin{figure*}
\includegraphics[scale=0.5]{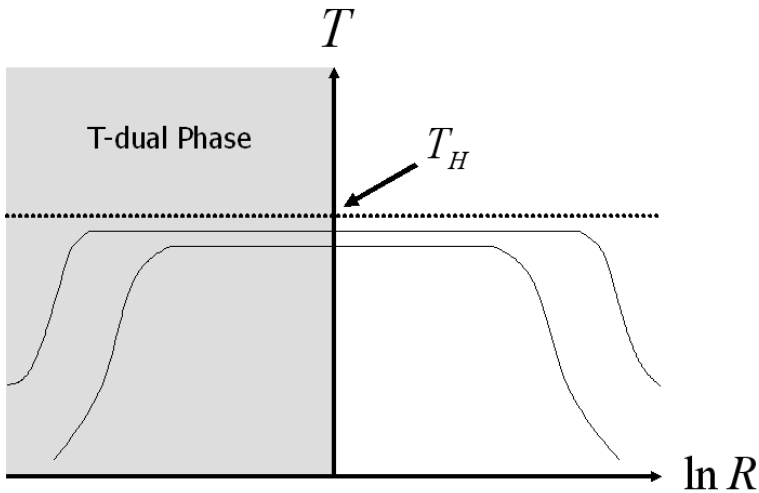}
\caption{Temperature (vertical axis) of a gas of heterotic superstrings as a function of box radius
(in units of the string length). $T_H$ is the Hagedorn temperature, the maximal temperature which
the gas of string can achieve. How close to $T_H$ the temperature $T$ actually gets depends on
the total entropy of the system. Shown are two curves corresponding to different entropies (the top
curve has a higher entropy).}
\end{figure*}

To obtain a cosmological scenario we need not only kinematics of
string gas cosmology, but also dynamics. At the present time we
do not have a first principles dynamics which comes from string theory.
Neither Einstein gravity nor dilaton gravity can be applied in the
Hagedorn phase since these frameworks are not consistent with
the symmetries of string theory which are expected to be present in
the Hagedorn phase. There are two possible dynamical scenarios.
In the first, the universe starts in a quasi-static Hagedorn phase with
all spatial dimensions wrapped by strings. The decay of string winding
modes into loops triggers the dynamical breaking of the T-duality
symmetry of the string gas and the transition to the radiation stage
of Standard Big Bang cosmology. Figure 2 is a sketch of the time
evolution of the cosmological scale factor $a(t)$ according to this
scenario \cite{BV}.

\begin{figure*}
\includegraphics[scale=0.5]{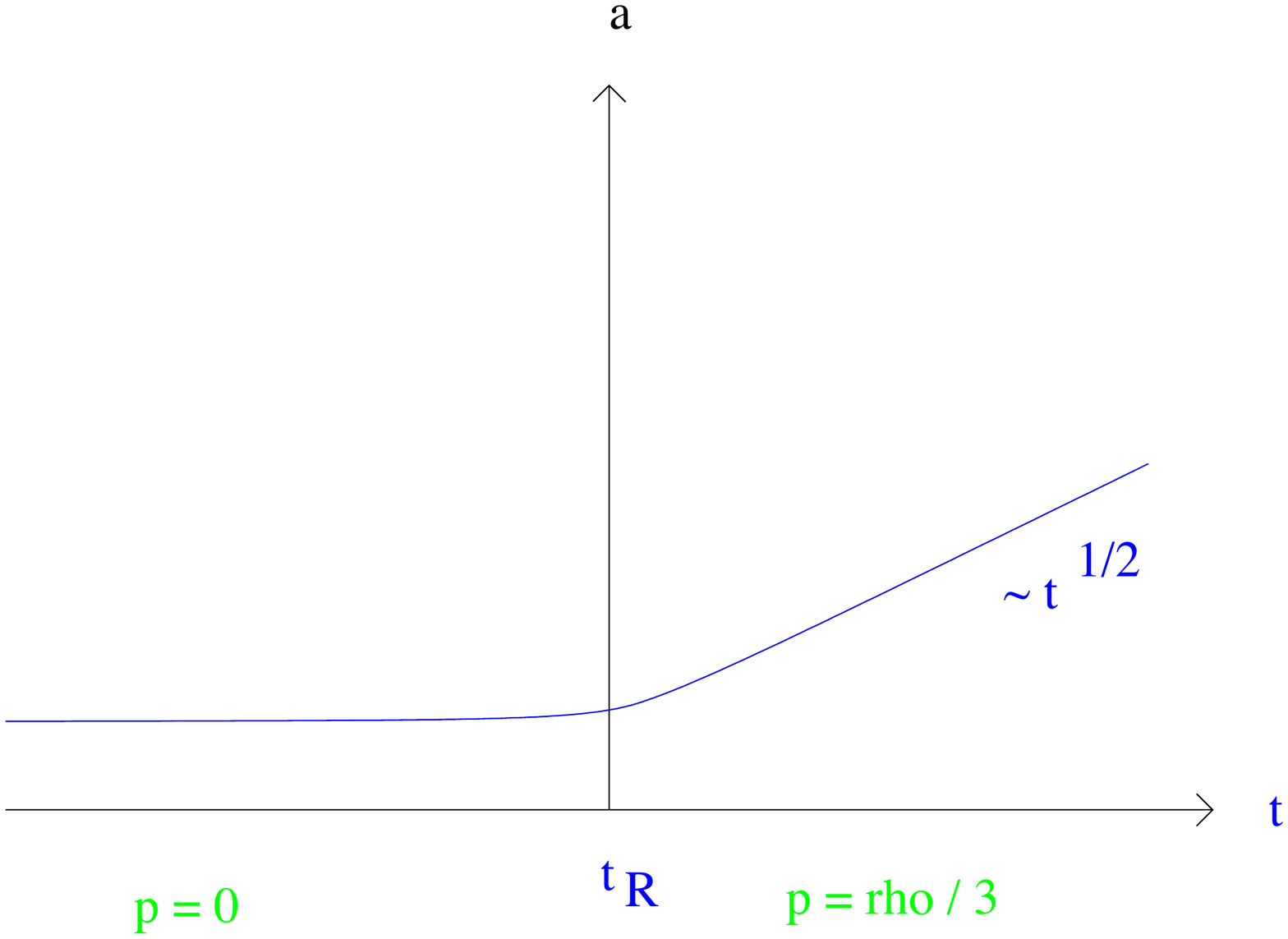}
\caption{Evolution of the scale factor in string gas cosmology. For $t < t_R$ we
are in the Hagedorn phase. At $t_R$ a phase transition to the radiation phase of
Standard Cosmology takes place.}
\end{figure*}

An alternative scenario is that the universe undergoes a cosmological bounce
in which $R$ starts out at the far left of Figure 1, i.e. with $R \ll l_s$, where $l_s$
is the string length, passes through the Hagedorn phase, and then enters the
usual radiation phase of Standard Cosmology. In terms of the light variables,
the phase with $R \ll l_s$ and $R$ increasing corresponds to a contracting
phase. Thus, this second scenario corresponds to a cosmological bounce
mediated by a gas of strings \cite{Biswas}. However, in the following we will 
explore the first possibility, namely that the universe starts
in a long quasi-static Hagedorn phase.

String theory is mathematically well-defined in ten space-time dimensions.
The idea of string gas cosmology is to start in a thermal state in which
all spatial dimensions are equivalent, and to then explain why only three
spatial dimensions effectively de-compactify \cite{BV}, as opposed to the usual
approach in string motivated field theory cosmology where one assumes
from the outset that our three spatial dimensions are special and one
then requires some ad-hoc compactification mechanism. Thus, we start
with a thermal gas of strings in which the momentum and winding modes
about all spatial dimensions are excited. The presence of winding modes
prevents spatial sections from expanding. The only way that a spatial
dimension can become large is if the winding modes about that dimension
can decay. Decay of winding modes, however, requires winding mode
interactions. Since such interactions require the string world sheets to
intersect, the interaction probability is negligible in more than
three spatial dimensions, as long as there are no long-range forces
between the strings \cite{BV} (see also \cite{Mairi} for a numerical
study). The three dimensions in which winding modes can annihilate
may not all start to expand at exactly the same time, but, as long as some
string winding modes are still present there is an isotropization
mechanism which is at work \cite{Scott1}.

In the string gas cosmology setup the six spatial dimensions in which
the winding modes were not able to annihilate are confined by the
gas of winding and momentum modes to remain at the string scale
\cite{Scott2, Subodh1}. Thus, the size moduli of string theory are
naturally stabilized in string gas cosmology. This corresponds to
moduli stabilization at enhanced symmetry points \cite{Scott3, Alex}.
In the case of heterotic superstring theory it can be shown
explicitly \cite{Subodh2} that this moduli trapping mechanism is
consistent with late time cosmology. The presence of winding
modes can also trap shape moduli of the internal dimensions, as
was shown in \cite{Edna}. The one modulus which is not stabilized
by intrinsic stringy effects is the dilaton. The dilaton, however, can
be stabilized by invoking gaugino condensation \cite{Frey}, without
destabilizing the size moduli. Gaugino condensation then leads
to (typically high scale) supersymmetry breaking \cite{Wei}. The
bottom line is that moduli stabilization, the Achilles heel of many
other approaches to string cosmology, appears to be in good
controle in the context of string gas cosmology.

\section{String Gas Cosmology, Structure Formation and the Planck Results}

In inflationary cosmology and some of its alternatives, the source of
cosmological perturbations is quantum vacuum fluctuations \cite{Mukh}.
A justification of this idea is as follows: the exponential expansion of
space dilutes the density of any excitations which may have been present
at the beginning of the inflationary phase, leaving behind a vacuum state
of matter. In contrast, in string gas cosmology the initial state is a hot gas
of strings in thermal equilibrium at a temperature close to $T_H$, the
Hagedorn temperature. Hence, in string gas cosmology the source
of inhomogeneities is thermal fluctuations of a gas of strings. As
realized in \cite{NBV}, this leads to an almost scale-invariant spectrum
of curvature perturbations at late times, and similarly \cite{BNPV2} to
an almost scale-invariant spectrum of gravitational waves. The 
spectrum of cosmological perturbations has a small red tilt, like in the
case of inflation. However, the spectrum of gravitational waves has
a small blue tilt \cite{BNPV2}, unlike in the case of inflation where a
red tilt is inevitable (provided matter is used which obeys the usual energy
conditions).

Cosmological fluctuations should be viewed as a superposition of
small amplitude plane wave inhomogenities. If we expand the full
equations of motion for space-time and matter to linear order
in these fluctuations, each wave will evolve independently. In inflationary
cosmology, each wave corresponds to a harmonic oscillator which
begins in its vacuum state on sub-Hubble scales and is stretched by
the accelerated expansion of space to super-Hubble lengths where
the wave function is squeezed and classicalizes via the intrinsic
nonlinearities of the system \cite{KPS, Martineau}. We now compare 
this setup to what happens in string gas cosmology.

\begin{figure*}
\includegraphics[scale=0.5]{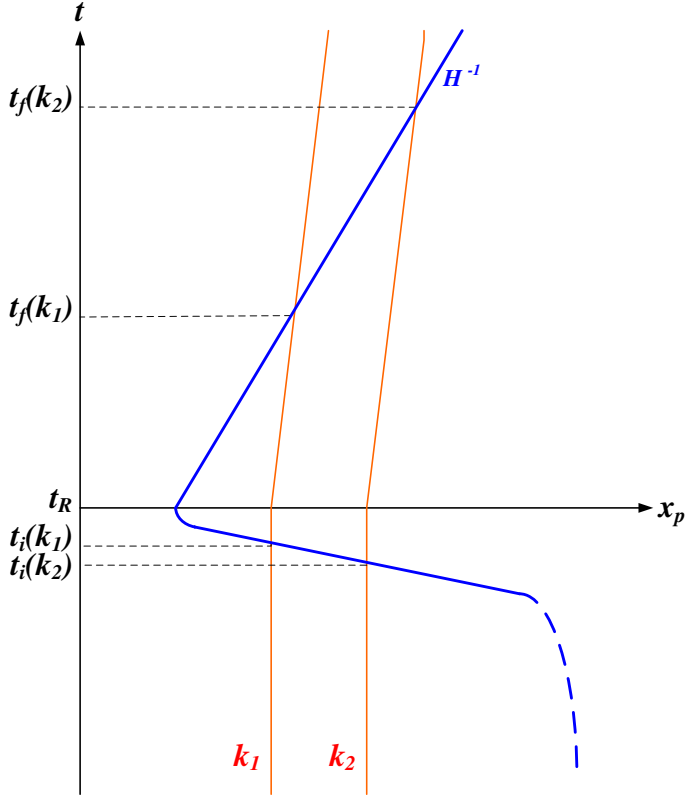}
\caption{Space-time sketch of string gas cosmology. The vertical axis
is time, with $t_R$ denoting the end of the Hagedorn phase. The horizonal
axis is physical distance. The blue curve which increases linearly for $t  > t_R$
is the Hubble radius which is infinite deep in the Hagedorn phase. The
red curves labelled by $k_1, k_2$ indicate the wavelength of fluctuation modes
which have constant comoving length. Since the Hagedorn phase is
static, these curves are vertical during this phase. The times $t_i(k)$ 
and $t_f(k)$ are the instances when the length of the wave $k$ crosses
the Hubble radius at the end of the Hagedorn phase and at late times,
respectively.}
\end{figure*}

A space-time diagram of string gas cosmology is shown in Fig. 3. The
vertical axis is time, the horizontal axis indicates physical spatial length.
The Hagedorn phase corresponds to times earlier than $t_R$. During
the Hagedorn phase space is static and hence the Hubble radius is
infinite. After the decay of the string winding modes our three dimensional
space starts to expand according to the usual laws of Standard Big Bang
cosmology. The Hubble radius drops to a microscopic value at $t_R$ and
then expands linearly as shown in the blue curve. The physical wavelength
of fluctuation modes in constant in the Hagedorn phase and then increases
in proportion to the scale factor $a(t)$ after $t_R$.

We first compare the ways in which inflation and string gas cosmology,
respectively, solve the horizon problem of Standard Big Bang cosmology
and lead to the possibility of a causal structure formation mechanism.
In inflationary cosmology it is the accelerated expansion of space which
renders the horizon much larger than the Hubble radius and ensures that
the past light cone of our current observer fits into the horizon at $t_{rec}$,
the time of recombination. In string gas cosmology a long Hagedorn phase
will similarly allow the horizon to become much larger than the Hubble
radius at $t_R$. In inflationary cosmology, it is again the accelerated 
expansion of space which allows fluctuation modes which are currently
observed on large scales to be pushed far outside of the Hubble radius.
In string gas cosmology, the wavelengths of the perturbation modes is
constant during the Hagedorn phase, but the Hubble radius decreases
dramatically such that modes become super-Hubble at the end of the
Hagedorn phase. In both cases, fluctuations not only have a wavelength
smaller than the horizon, but also smaller than the Hubble radius. This
enables a causal generation mechanism.

Assuming that the string scale is close to the scale of particle
physics Grand Unification, which is the preferred value for heterotic
superstring particle phenomenology \cite{GSW}, the physical
wavelength of fluctuation modes which are observed today
is of the order $1 {\rm mm}$. While this scale seems microscopic
from the point of view of cosmology, it is very large compared to
the string scale or the Planck scale. In inflationary cosmology,
the wavelength of these fluctuations is exponentially smaller than
this scale at the beginning of the period of inflation, thus leading
to the ``trans-Planckian problem'' for fluctuations \cite{Jerome}:
it is not justified to use Einstein gravity and low energy effective
classical matter physics to study the origin and early evolution
of fluctuations. In contrast, in string gas cosmology the fluctuation
modes of interest to us are safely in the far infrared for all times,
and thus safe from the trans-Planckian problem.

In contrast to the case of inflationary cosmology, in string gas cosmology
the inhomogeneities are not vacuum fluctuations, but rather thermal fluctuations.
Importantly, they are not thermal fluctuations of a gas of point particles,
but of a gas of fundamental strings. Hence, the thermal fluctuations are
described by string thermodynamics (see e.g. \cite{Nivedita}). The
computation of the spectrum of cosmological fluctuations and gravitational
waves now proceeds as follows \cite{NBV, BNPV2}: we first compute
the matter fluctuations in the Hagedorn phase, using relations of
string thermodynamics \cite{Nivedita}. In a second step, we use
the Einstein constraint equations to relate the matter fluctuations
to metric fluctuations. This is done mode by mode at the time when
the wavelength crosses the Hubble radius. Finally, we evolve the
metric perturbations on super-Hubble scales until they re-enter the
Hubble radius at late times using the equations of the theory of
cosmological perturbations (see e.g. \cite{MFB, RHBfluctrev} for
reviews).

Our method relies on three key assumptions: firstly the existence
of an initial quasi-static phase containing a thermal gas of strings.
Secondly, we posit a fast transition from the Hagedorn phase
to the radiation phase of Standard cosmology.
Thirdly, we assume the validity of Einstein gravity in the far infrared,
an assumption which we use both in converting matter fluctuations
to metric ones, and in evolving the perturbations to late times.

We begin with the ansatz for a space-time metric containing
both linear cosmological fluctuations (also called ``scalar metric
fluctuations'') and gravitational waves (``tensor metric fluctuations''):
\begin{equation}
d s^2 \, = \, a^2(\eta) \bigl( (1 + 2 \Phi(x, \eta) )d\eta^2 - 
[ (1 - 2 \Phi)\delta_{ij} + h_{ij} ]d x^i d x^j\bigr) \,
\end{equation}
where $\eta$ is conformal time, $a(\eta)$ is the scale factor
describing the background cosmology, $\Phi(x, \eta)$ are
the cosmological perturbations which depend on the
spatial coordinates $x$ and on time, and the transverse traceless
tensor $h_{ij}(x, \eta)$ describes the gravitational waves
(see \cite{MFB, RHBfluctrev}). We have chosen a gauge
(coordinate system) in which the metric corresponding to
the cosmological perturbations is diagonal, and assumed
that there is no anisotropic stress (which leads to the fact
that there is only one non-trivial function $\Phi$ characterizing
these fluctuations. Note that Latin indices represent spatial
coordinates.

The Einstein constraint equations determine the scalar
and tensor metric fluctuations in terms of the fluctuations
of the energy-momentum tensor. Specifically,
 \begin{equation}  \label{scalarexp}
\langle |\Phi(k)|^2\rangle \, = \, 16 \pi^2 G^2 
k^{-4} \langle\delta T^0{}_0(k) \delta T^0{}_0(k)\rangle \, , 
\end{equation}
and
\begin{equation} \label{tensorexp} 
\langle |h(k)|^2\rangle \, = \, 16 \pi^2 G^2 
k^{-4} \langle\delta T^i{}_j(k) \delta T^i{}_j(k)\rangle \, ,
\end{equation} 
where $G$ is Newton's gravitational constant,
where in the last equation $h$ is the amplitude of
each of the two polarization modes of gravitational
waves, and on the right hand side an average of
the off-diagonal spatial matrix elements (i.e. $i \neq j$)
is implicit. The expectation values on the right hand
side of the above equations indicate thermal
expectation values.

Since the wavelengths of the modes we are interested
in are always in the far infrared compared to the string
scale, we can safely use the perturbed Einstein equations
to study the evolution of the fluctuations. From the
theory of cosmological fluctuations we know that, since
the equation of state parameter $1 + w$ (where $w$ is
the ratio of pressure $p$ to energy density $\rho$)
does not change by more than a factor of order one
during the transition from the Hagedorn phase to the
Standard Cosmology phase (in contrast to what
happens in inflationary cosmology during the transition
between the inflationary phase and the post-inflation
period), both $\Phi$ and $h$ remain
constant while on super-Hubble scales. Hence, it
will be the values of $\Phi$ and $h$ computed at
Hubble radius crossing at the end of the Hagedorn phase
which are relevant for current observations.

Let us now turn to the determination of the initial fluctuations.
In general, for thermal fluctuations the energy density
perturbations are determined by the specific heat capacity
$C_V$ (in a volume $V$), and by the temperature $T$.
If $C_V$ is the specific heat capacity in a volume of
radius $R$, the resulting energy density variations are
given by
\begin{equation} \label{drho}
\langle \delta\rho^2 \rangle \,  = \,  \frac{T^2}{R^6} C_V \, . 
\end{equation}
What is special about string thermodynamics is that
for strings in a compact space of radius $R$, the specific
heat capacity has holographic scaling with $R$ \cite{Nivedita, Ali}
\begin{equation} \label{specheat}
C_V  \, \approx \, 2 \frac{R^2/l_s^3}{T \left(1 - T/T_H\right)}\, .,
\end{equation} 
where $l_s$ is the string length.

By combining these equations we can compute the
power spectrum of cosmological fluctuations which is
defined as
\be \label{power2} 
P_{\Phi}(k) \,  \equiv  \, {1 \over {2 \pi^2}} k^3 |\Phi(k)|^2 \, .
\ee
Inserting the value of $\Phi(k)$ at Hubble radius crossing in
terms of the density fluctuations from (\ref{scalarexp}), making
use its representation (\ref{drho}) in terms of the specific
heat capacity, and then making the replacement (\ref{specheat})
we obtain the following expression for the power spectrum in
terms of the temperature $T(k)$ when the mode $k$ exits
the Hubble radius:
\be \label{sresult}
P_{\Phi}(k) \, = \,  \left(\frac{l_{pl}}{l_s}\right)^4{T(k) \over {T_H}} {1 \over {1 - T(k)/T_H}} 
\,  , ,
\ee 
where $l_{pl}$ is the Planck length (determined by $G$).

As can be seen from Figure 1, to first approximation the temperature
$T(k)$ is independent of $k$. To next order, however, we notice that
$T(k)$ in a slightly decreasing function of $k$ as $k$ increases, since
the temperature starts to decrease as we near the exit from the Hagedorn
phase. Thus, string gas cosmology, like inflation, predicts a roughly
scale-invariant spectrum of cosmological fluctuations with a slight red
tilt \cite{NBV}. The spectral tilt $n_s - 1$ is given by \cite{BNP}
\be
\label{scalartilt}
n_s-1 \, = \,  (1- \frac{T(k)}{T_H})^{-1} k \frac{dT(k)/ T_H}{d k},
\ee
which is negative since $T(k)$ is a decreasing function of $k$,

The above computation contains two parameters which from our point
of view are free, the first being the ratio of the string length to the Planck
length, and the second the deviation of the temperature from $T_H$ during
the Hagedorn phase. The latter depends on the total entropy of the
system \cite{BV}, the former is set by the string model. Making use
of the value of the string length preferred for particle physics reasons
in \cite{GSW}, and taking the ratio of temperatures on the right hand side
of (\ref{sresult}) to be of the order one, we obtain an amplitude of the
spectrum which is consistent with observations. However, from the
effective theory analysis which we have presented, we must take
$l_s$ and $1 - T(k_0)/T_H$ as free parameters (where $k_0$ is the
pivot scale). These parameters can be fixed by demanding agreement
with the current CMB data. Once these parameters are fixed, however,
both the amplitude and spectral tilt of the gravitational wave
spectrum are determined.

As indicated in (\ref{tensorexp}), the gravitational wave spectrum
is given by the off-diagonal pressure fluctuations. String thermodynamics
allows the computation of all stree-energy tensor correlation functions.
The result for the  off-diagonal stress correlation function is \cite{Ali}
\begin{equation}
<|T_{ij}(R)|^2> \, \sim \, {T \over {l_s^3 R^4}} (1 - T/T_H) 
\ln^2{\left[\frac{1}{l_s^3 T R^{-2}}(1 - T(k)/T_H)\right]}\, .
\end{equation}
Note that the factor $(1 - T/T_H)$ is in the numerator instead
of in the denominator as in the case of the energy density
correlation function. Inserting this expression into (\ref{tensorexp})
yields the following result for the power spectrum of gravitational
waves
\be \label{tresult} 
P_h(k) \, \sim \,
\left(\frac{l_{pl}}{l_s}\right)^4 \frac{T(k)}{T_H}(1 -
T(k)/T_H)\ln^2{\left[\frac{1}{l_s^2 k^2}(1 - T(k)/T_H)\right]} \, .
\ee
This corresponds to a scale-invariant spectrum, but this time with
a blue tilt. The tensor tilt $n_t$ is
\be \label{cons1}
n_t \, = \,  \frac{1 - 2 T(k)}{1- T(k) / T_H} k\, \frac{d T(k) / T_H}{d k} 
\, = \,  -(n_s-1)(2\frac{T(k)}{T_H} - 1) \, .
\ee
To first approximation, the magnitude of the blue tilt of the tensor
spectrum equals the magnitude of the red tilt of the scalar spectrum.

The difference in the sign of the tilt of the gravitational wave spectrum
is the key and most rebust criterion which differentiates between 
cosmological inflation and string gas cosmology \cite{BNPV2, BNP}. The reason why the
tilt in inflation is negative (i.e. a red spectrum) is that the amplitude
of the gravitational wave spectrum is set by the Hubble constant $H$,
and that during inflation $H(t)$ is a decreasing function of time (as
long as matter satisfies the ``null energy condition''). Thus, long wavelength
fluctuations exit the Hubble radius at a larger value of $H$ and thus
with a larger amplitude of gravitational waves. On the other hand,
in string gas cosmology the gravitational wave spectrum is determined
in terms of the off-diagonal pressure fluctuations which are proportional
to the pressure. Earlier in the Hagedorn phase the pressure is closer
to zero, and hence large-scale modes exit the Hubble radius with a
smaller amplitude of the pressure perturbations than small wavelength
modes, and a blue spectrum results. Since the temperature $T(k)$ in
the Hagedorn phase is close to $T_H$, the consistency relations
approximately reads
\be
n_t \, \simeq - (n_s - 1) \, .
\ee
Based on the Planck data \cite{Planck}, string gas cosmology thus predicts
\be
n_t \, = \, 0.03 \pm 0.01 \, .
\ee

The tensor to scalar ratio $r$ is also a prediction of string gas cosmology.
By combining (\ref{sresult}) and (\ref{tresult}) we obtain
 \be \label{cons2}
r(k) \, = \, (1 - \frac{T(k)}{T_H})^2 \ln^2{\left[\frac{1}{l_s^2 k^2}(1 - \frac{T(k)}{T_H})\right]} \, .
\ee
Note that, as in the case of inflationary cosmology, the value of $r$ depends on
the scale $k$.                     

At this stage, the background of string gas cosmology is not sufficiently developed
to be able to make a specific prediction for the tensor to scalar ratio. From
(\ref{sresult}) and (\ref{scalartilt}) we see that the amplitude and tilt of the
scalar spectrum depend on the ratio $l_{pl} / l_s$, the factor $1 - T/T_H$
and on the $dT(k) / dk$. The last factor is not known since we do not have
an analytical description of the exit from the Hagedorn phase. Just considering
the factors $(1 - T/T_H)$ we would expect
\be
r \, \sim \, (1 - T/T_H)^2 
\ee
which is expected to be significantly smaller than $1$. It would be interesting
to model the exit from the Hagedorn phase in order to obtain an actual
prediction for $r$.

The Planck and joint Planck/BICEP2 results for $r$ yield a bound of $r < 0.1$.
Thus, the results are completely consistent with the predictions of string
gas cosmology. To differentiate inflation and string gas cosmology it will be
crucial to determine the tensor tilt. The original BICEP2 results \cite{BICEP2}
favored a blue tilt of the tensor spectrum  \cite{YiWang}, 
and hence favored string gas cosmology over inflation as was stressed in
\cite{BNP}. 

The experimental prospects for measuring the tensor tilt depend on the
amplitude $r$. For a value $r = 0.05$, a careful analysis of B-mode polarization
data including de-lensing will allow an identification of a tensor tilt variance of
$\sigma(n_t) = 0.04$ \cite{Simard}, very close to the prediction of string gas
cosmology. The current upper bound on the tensor tilt are
\be
n_t \, < \, 0.15 \,
\ee
(making use of constraints from pulsar timing, direct detection experiments
and nucleosynthesis \cite{Stewart}).

Since the fluctuations in string gas cosmology are of thermal origin, 
thermal non-Gaussianities will be produced. However, these
non-Gaussianities are Poisson suppressed on scales larger than
the characteristic scale of the thermal fluctuations, the inverse
temperature. Since the temperature in the Hagedorn phase is close 
to the string scale, the non-Gaussianities on observable scales
will be highly suppressed \cite{YiWang2}. Hence, observing non-Gaussianities
on cosmological scales would be a serious challenge for string gas
cosmology. On the other hand, the Planck satellite has not seen
any non-Gaussianities, and hence also in this respect the predictions
of string gas cosmology are consistent with current observations.

There is one type of non-Gaussianities which could be present
in string gas cosmology: if we are dealing with a string theory in which
cosmic superstrings \cite{Witten} are stable (see \cite{Myers} for
a discussion of the criteria for this to be the case), then string gas
cosmology would leave behind a network of cosmic strings in our
three dimensional space. As studied in detail in the case of cosmic
strings (see e.g. \cite{CSreviews} for reviews on cosmic strings
and cosmology), the network of cosmic superstrings would take
on a ``scaling solution'' in which the network of strings looks identical
at all times when all lengths are scaled to the Hubble radius.
The scaling solution corresponds to a fixed number $N$ of infinite
string segments crossing each Hubble volume, and a distribution
of string loops with radius smaller than $t$ which are produced
by the interactions between the infinite strings. As a consequence
of the energy which is trapped in the strings,  cosmic superstrings 
(like cosmic strings) leave behind clear signatures in cosmological
observations. The power spectrum of the string-induced fluctuations
is approximately scale-invariant (see e.g. \cite{Turok}). The non-Gaussianities
are prominent in position space maps: line discontinuities in CMB
temperature maps \cite{KS}, rectangles in the sky with direct
B-mode polarization \cite{Holder1}, and thin wedges in 21cm redshift
maps (extended in the sky over degree scale but thin in redshift
direction) \cite{Holder2}. The current bound on the string tension $\mu$
from not detecting any string-specific signatures is 
\be
G \mu \, < \, 2 \times 10^{-7} \, .
\ee
This bound comes from combining Planck data with that of smaller
angular scale telescopes \cite{Dvorkin, PlanckTD}. With a dedicated
position space search using Planck data, a reduction of this limit 
might be possible. A study of the potential of the South Pole Telescope
to contrain $G \mu$ indicated \cite{Danos} that an improvement of
the bound by one order of magnitude should be possible (for a recent
discussion of signals of cosmic strings in new observational windows
the reader is referred to \cite{RHBCSrev}).

\section{Discussion and Conclusions}

Assuming the cosmological background given in Fig. 1, string gas cosmology
naturally solves the horizon problem of the Standard Big Bang model. In
contrast to inflationary cosmology, the string gas cosmology does not provide
a mechanism to produce spatial flatness, and it also assumes a large
initial size and entropy of space. As shown above, string gas cosmology
leads to a structure formation scenario which yields predictions which are
in good agreement with all current observations, and makes predictions
for future observations with which the model can be distinguished from
cosmological inflation.

The Achilles heel of string gas cosmology is the fact that at the current
time we do not have a mathematical desciption of the Hagedorn phase.
In this phase, non-perturbative string theory will be crucial. Einstein and
dilaton gravity are not applicable since these background actions are
not consistent with the key symmetries which distinguish string theory
from point particle theory. A possible framework to study the dynamics
of the Hagedorn phase is ``double field theory'' \cite{Hull}, a setup in which
the number of dimensions is doubled, the two copies being related by
T-duality. 

\ack{The author thanks D. Wiltshire and F. Bouchet for the invitation
to prepare this contribution. The author's research is supported 
by an NSERC Discovery Grant, and by funds from the Canada 
Research Chair program.}

\end{document}